\documentstyle[aps,twocolumn,psfig]{revtex}

\begin{document}
\tighten
\title{\bf Critical properties of the topological Ginzburg-Landau 
model}

\author{C. de Calan$^{1}$, A. P. C. Malbouisson$^{2}$,
F. S. Nogueira$^{1}$, N. F. Svaiter$^{2}$}
\address{$^{1}$Centre de Physique Th\'eorique, Ecole Polytechnique, 91128 
Palaiseau, FRANCE\\ 
$^{2}$Centro Brasileiro de Pesquisas Fisicas - CBPF,
Rua Dr. Xavier Sigaud 150, Rio de Janeiro, RJ 22290-180,
BRAZIL}

\date{Received \today}

\maketitle

\begin{abstract}
We consider a Ginzburg-Landau model for superconductivity with a Chern-Simons 
term added. The flow diagram contains two charged fixed points 
corresponding to the tricritical and infrared stable fixed points. The 
topological coupling controls the fixed point structure and eventually 
the region of first order transitions disappears. We compute the critical 
exponents as a function of the topological coupling. We obtain 
that the value of the $\nu$ exponent does not vary very much from the 
$XY$ value, $\nu_{XY}=0.67$. This shows that the Chern-Simons term does 
not affect considerably the $XY$ scaling of superconductors.  
We discuss briefly the possible 
phenomenological applications of this model.

Pacs numbers: 74.20De, 11.10Hi.
\end{abstract}
\pacs{ }

\section{Introduction}
The Ginzburg-Landau model (GL) has been introduced almost half a century 
ago \cite{Ginzburg} as a phenomenological model for superfluidity and 
superconductivity. In its one component order parameter version it has been 
used with remarkable success as a statistical mechanics model for the 
critical phenomena of systems lying in the same universality class as the 
Ising model \cite{ZJ}. In its $N$ component version coupled to 
Abelian gauge fields it has been used as a model for superconductivity and 
liquid crystals \cite{deGennes}. However, in this last situation the 
$\epsilon$-expansion, which works very well in the non-gauged version, 
seems to be insufficient to describe unambiguously the critical properties 
of the model \cite{HLM,Hikami}. Only in the large $N$ limit the 
$\epsilon$-expansion gives consistent results \cite{HLM}. The trouble is the 
absence of a second order (infrared stable) fixed point in the flow 
diagram. This result is physically correct only in the extreme type I 
regime $u/e^{2}<<1$ (here $u$ is the quartic scalar self-coupling and 
$e$ is the charge), 
where we expect a weak first order phase transition. This 
regime is also well described by the fluctuation corrected mean field 
analysis of Halperin {\it et al.} \cite{HLM}. The weak first order transition 
has been probed experimentally in certain classes of liquid crystals, where 
essentially the same GL model holds \cite{exp}. For the extreme 
type II region ($u/e^{2}>>1$), a second order fixed point is expected. Indeed, 
this result follows from numerical studies performed in a lattice dual 
GL model \cite{Dasgupta}. Therefore, the prediction of a first order 
transition even in the type II regime seems to be an artifact of the 
$\epsilon$-expansion. More recent works support also this point of view 
\cite{KleinertI,Berg,Herbut,HerbutI,Radz}. Also, it has been shown that the 
renormalization group (RG) in a fixed dimension approach is more appropriate 
\cite{KleinertI,Berg,Herbut}. The flow diagram in the $u-e^{2}$ plane 
exhibits in general four fixed points: Gaussian and superfluid (or XY), both 
uncharged, and the tricritical and superconducting which are charged fixed 
points. The Gaussian fixed point is trivial and describes the mean field 
critical behavior of a $O(2)$ model (we shall work with $N=2$). The 
superfluid fixed point describes the $\lambda$-transition in $He^{4}$ and 
lies in the same class of universality as the XY model. The tricritical 
fixed point is over a line, called tricritical line, which separates the 
regions of first and second order phase transition. This fixed point is 
attractive along the tricritical line but repulsive in the direction nearly 
parallel to the $u$-axis. Finally, the superconducting fixed point is 
a charged infrared stable fixed point and describes a second order 
superconducting phase transition. Bergerhoff {\it et al.} 
\cite{Berg} obtained this 
flow diagram using Wilson's RG \cite{Wilson} in a non-perturbative version 
called {\it exact renormalization group} \cite{Wegner}. They used the 
background field formalism to control gauge invariance which is in principle 
violated due to the presence of the cutoff (further discussions on this 
subtle point in Wilson's RG can be found in refs.\cite{Reuter}). This flow 
diagram has been also obtained by Herbut and Tesanovic  
\cite{Herbut} using a simpler method. They performed a 1-loop calculation in a fixed dimension approach.  

Recently a flow diagram with qualitatively the same structure has been also 
obtained for a GL model with a topological Chern-Simons (CS) term by 
Malbouisson {\it et al.} \cite{Malbouisson}. Their analysis was performed 
using Wilson's RG in perturbative form, which is not manifestly gauge 
invariant due to the cutoffed integrals. Due to the presence of a CS term, an 
intrinsically three dimensional object, their calculations were performed in 
$d=3$ and $N=2$, resulting therefore in an uncontrolled approximation since 
there is no small parameter as $\epsilon$ or $1/N$. The same type of model
has been considered earlier by Kleinert and Schakel \cite{FUp} using a 
different scaling. They performed a 1-loop calculation of critical 
exponents. However, their scaling did not allows for a consistent zero CS mass 
limit since Feynman graphs were evaluated at zero external momenta. The 
reason comes from the graph shown in Fig.1. When evaluated at zero external 
momenta this graph gives zero due to the structure of the CS gauge field 
propagator. On the other hand, in the same scaling this graph is infrared 
divergent if no CS mass is included in the action. Thus, it is not 
legitimate to perform the zero CS mass limit in the scaling considered 
in \cite{FUp} as observed by the authors themselves.  
There are many other  
RG studies of bosonic CS models in the literature but the $F^{2}$ term is 
almost always absent \cite{Park}. The presence of such a term is crucial 
in order to obtain the flow diagram of ref. \cite{Malbouisson}. Moreover, it 
is desirable to recover the usual GL model in the limit of zero CS mass.  

In this paper we consider further the topological model of refs.
\cite{Malbouisson,FUp}, performing the calculation at the critical point 
\cite{ZJ}. We consider two different approximations. In a first step 
(section II) we perform a 1-loop calculation of the RG functions assuming 
that the same scale holds for both the order parameter and 
the gauge field. In this context we find the following main features: 
(1) For the CS coupling smaller than a certain critical value there are no 
charged fixed points; (2) There exists a interval of CS couplings such that 
two charged fixed points are found, corresponding respectively to the 
tricritical and second order (infrared stable) fixed points. In this interval 
it is possible to find respectable values for the $\nu$ exponent but not 
for the $\eta$ exponent, eventually violating the scaling relations; (3) 
For larger CS coupling, outside the interval mentioned in (2), the region 
of first order behavior is lost since the tricritical fixed point assumes 
an unphysical value. The second step (section III) consists in improving 
upon the 1-loop result of  section II by distinguishing the scales of the 
order parameter and the gauge field. In this approximation we follow an 
idea of Herbut and Tesanovic in order to relate both scales and obtain in 
this way the RG flow. In this case we have the tricritical and second order 
fixed points even in the limit of zero CS coupling, consistent with the 
fixed point structure of the conventional GL model. However, once again the 
tricritical fixed point becomes unphysical for the CS coupling larger than 
a certain critical value. In this case we obtain that 
all the critical exponents have respectable physical values. Moreover, 
the $\nu$ exponent as a function of the topological coupling does not 
deviate very much from $XY$ scaling.
Finally, in section IV we discuss our results and 
the possible applications.  

\section{Model and RG results}

Our starting point is the following action:

\begin{eqnarray}
S&=&\int d^{3}x\left[|(\nabla-ie_{0}\vec{A}_{0})\psi_{0}|^{2}
+r_{0}|\psi_{0}|^{2}+\frac{u_{0}}{2}|\psi_{0}|^{4}\right.\nonumber\\
&+&\left.\frac{1}{8\pi\mu_{0}}(\nabla\times\vec{A}_{0})^{2}+
i\frac{\theta_{0}}{2}
\vec{A}_{0}\cdot(\nabla\times\vec{A}_{0})\right]+S_{gf},
\end{eqnarray}
where the subindex $0$ denotes bare quantities. The above action is a 
standard GL model with a CS term added. $S_{gf}$ is the gauge fixing term 
which is given by

\begin{equation}
S_{gf}=\int d^{3}x\frac{1}{2\alpha_{0}}(\nabla\cdot\vec{A}_{0})^2,
\end{equation}
with $\alpha_{0}$ being the bare gauge fixing parameter.
The bare propagator for the gauge field is given by

\begin{eqnarray}
D_{ij}(k)&=&\frac{4\pi\mu_{0}}{k^{2}+g^{2}_{0}}\left[\delta_{ij}-\frac{
k_{i}k_{j}}{k^{2}}-g_{0}\frac{\epsilon_{ijk}k_{k}}{k^{2}}\right.\nonumber\\
&+&\left.\frac{\alpha_{0}}{4\pi\mu_{0}}\left(1+\frac{g_{0}^2}
{k^2}\right)\frac{k_{i}k_{j}}{
k^2}\right],
\end{eqnarray}
where $g_{0}=4\pi\mu_{0}\theta_{0}$.

Now we write the renormalized action as a sum of $S'+\delta S$ where $S'$ is 
the same as $S$ with the bare quantities replaced by renormalized ones 
(in our notation it corresponds to drop the zeroes). $\delta S$ is the 
counterterm part which is given by

\begin{eqnarray}
\delta S&=&\int d^{3}x\left[(Z_{\psi}-1)|(\nabla-ie\vec{A})\psi|^{2}
+(r_{0}Z_{\psi}-r)|\psi|^{2}\right.\nonumber\\
&+&\frac{u}{2}(Z_{u}-1)|\psi|^{4}
+\frac{Z_{\mu}-1}{8\pi\mu}(\nabla\times\vec{A})^{2}\nonumber\\
&+&\left.i\frac{\theta}{2}(Z_{\theta}-1)
\vec{A}\cdot(\nabla\times\vec{A})+\frac{Z_{\alpha}-1}{2\alpha}(\nabla\cdot
\vec{A})^2\right],
\end{eqnarray}
where the renormalized fields are given by $\vec{A}_{0}=\sqrt{Z_{A}}\vec{A}$ 
and $\psi_{0}=\sqrt{Z_{\psi}}\psi$. We shall perform our calculations at the 
critical point and therefore $r=0$. Also, we choose $r_{0}$ in such a way as 
to cancels the tadpole graphs.
The renormalization 
conditions for the renormalized vertex functions are given by 

\begin{eqnarray}
\Gamma^{(2)}_{\psi}(k)|_{k^2=0}&=&0,\\
\left.\frac{\partial\Gamma_{\psi}^{(2)}}{\partial k^2}(k)\right|_{k^2=p^2}
&=&1,\\
\left.\frac{1}{2}\frac{\partial}{\partial k^2}\left[P_{T}^{ij}(k)
\Gamma^{(2)}_{A,ji}\right]\right|_{k^2=p^2}=\frac{1}{4\pi\mu},\\
\left.\frac{1}{2}\frac{\partial}{\partial k^2}\left[P_{L}^{ij}(k)
\Gamma^{(2)}_{A,ji}\right]\right|_{k^2=p^2}=\frac{1}{\alpha},\\
\frac{\partial}{\partial k^2}\left(\epsilon_{ijk}k_{k}\Gamma^{(2)}_{A,ji}(k)
\right)\left.\right|_{k^2=p^2}&=&-2\theta,\\
\Gamma_{\psi}^{(4)}(k_{1},k_{2},k_{3},k_{4})|_{S.P.}&=&up,
\end{eqnarray}
where $u$ is a dimensionless coupling and 
S.P. denotes the symmetrical point which we take as being given by

\begin{equation}
k_{a}\cdot k_{b}=(4\delta_{ab}-1)\frac{p^{2}}{4}.
\end{equation}
$P^{ij}_{T}(k)$ and $P^{ij}_{L}(k)$ are the transverse and longitudinal 
projections given respectively by $P^{ij}_{T}(k)=\delta_{ij}-k_{i}k_{j}/k^2$ 
and $P^{ij}_{L}(k)=k_{i}k_{j}/k^2$.
Note that in the above renormalization conditions the gauge coupling and 
the scalar coupling are fixed at the same momentum scale.  
      
The renormalized couplings will depend on the momentum scale $p$ and 
the beta functions are defined through derivatives of the couplings with 
respect to $\log p$. Note that in order to preserve the $O(2)$   
symmetry of the four point function, we must consider the Feynman graphs with 
all the external momenta incoming at the vertices. If we proceed otherwise, 
that is, if we choose a convention which two external lines are incoming 
while the other two are outcoming, then the corresponding four point 
function is not symmetric when using the above symmetrical point. This is 
easily seen by performing a 1-loop calculation in the uncharged model with 
$O(N)$ symmetry. The resulting four point function 
$\Gamma_{\alpha\beta\gamma\delta}^{(4)}$ 
(the subindices are color indices) {\it is not} 
proportional to the $O(N)$ symmetric tensor 
$(\delta_{\alpha\beta}\delta_{\gamma\delta}+\delta_{\alpha\gamma}
\delta_{\beta\delta}+\delta_{\alpha\delta}\delta_{\beta\gamma})/3$ 
if the graphs are not evaluated with the convention that all the external 
momenta are incoming.

The Ward identies imply the exact relations $Z_{e}Z_{\psi}=1$, 
$e^{2}=e^2_{0}Z_{A}$ and $\alpha_{0}=\alpha Z_{A}$, where $Z_{e}$ is the 
charge renormalization. Let us define the following dimensionless gauge 
couplings, $\hat{e}^2=e^2/p$, $\hat{\theta}=\theta/p$, $f=4\pi\mu\hat{e}^2$ 
and $g=4\pi\mu\hat{\theta}$. 
The flow equations are given up to 1-loop order by
\begin{eqnarray}
\label{f}
p\frac{df}{dp}&=&-f+\frac{f^2}{16},\\
\label{g}
p\frac{dg}{dp}&=&\left(\frac{f}{16}-1\right)g,\\
\label{a}
p\frac{d\alpha}{dp}&=&-\eta_{A}\alpha,\\
\label{e}
p\frac{d\hat{e}^2}{dp}&=&(\eta_{A}-1)\hat{e}^2,\\
\label{t}
p\frac{d\hat{\theta}}{dp}&=&(\eta_{A}-1)\hat{\theta},\\
\label{u}
p\frac{du}{dp}&=&(2\eta_{\psi}-1)u+\frac{5}{8}u^{2}+\frac{\omega}{4\pi}f^{2},
\end{eqnarray}
where $\eta_{\psi}$ is the anomalous dimension for the scalar field defined 
by 

\begin{equation}
\eta_{\psi}=p\frac{d\log Z_{\psi}}{dp},
\end{equation}
while $\eta_{A}$ is the anomalous dimension for the gauge field which is 
defined by

\begin{equation}
\eta_{A}=p\frac{d\log Z_{A}}{dp}.
\end{equation}
$\omega$ is a function of $g$ which will be written later, together with the 
explicit expressions of the corresponding anomalous dimensions. Before 
doing this, let us discuss some important points concerning the general 
structure of the above flow equations. First, we note that $\eta_{A}$ does 
not appear in the flow equations for the couplings $f$, $g$ and $u$. This is 
because the flow equation for $\mu$ is given up to 1-loop order by

\begin{equation}
p\frac{d\mu}{dp}=-\eta_{A}\mu+\frac{f\mu}{16}.
\end{equation}
Thus, the flow diagram is completely determined by the couplings 
$f$, $g$ and $u$. This means that we do not need to know the expression 
of $\eta_{A}$. Another important point concerns the renormalization of 
$\theta$. It is a known fact in topological field theory that the CS mass 
does not renormalize at all orders in perturbation theory and for the 
model we are considering it can be verified by explicit calculation up 
to 2-loops \cite{Park}. Thus, $Z_{\theta}=1$ and $\theta=Z_{A}\theta_{0}$, 
which implies Eq.(\ref{t}). 

The most important point concerning the above 
flow equations is related to the gauge dependence. It can be shown that the 
beta function for the gauge couplings are gauge independent in a minimal 
subtraction scheme \cite{ZJ}. However, $\eta_{\psi}$ is gauge dependent. When 
$\eta_{\psi}$ is evaluated at the fixed point it gives the $\eta$ 
exponent. Note that exponents for the superconducting transition should be 
evaluated at the infrared stable charged fixed point, if we assume that it 
exists. At the superconducting fixed point we must have $\eta_{A}=1$. This 
means that the fixed point value of $\alpha$ must be $\alpha=0$, the 
Landau gauge. Since at the neighborhood of the superconducting fixed point 
$\alpha$ flows to the Landau gauge, critical exponents are evaluated for  
$\alpha=0$ 
and we shall fix it from the very beginning, as is customary in 
the literature.     

The explicit  analytical expressions of $\eta_{\psi}$ and 
$\omega$ are given by 

\begin{eqnarray}
\eta_{\psi}&=&-\frac{f}{4\pi}\left[\frac{3\pi}{4g^{2}}+\frac{\pi}{2}-\frac{
3\pi g^{2}}{4}+3g-\frac{3}{g}\right.\nonumber\\
&-&\left.\left(\frac{3}{2g^{2}}-1+\frac{3g^{2}}{2}\right)\arctan\left(
\frac{1-g^2}{2g}\right)\right],
\end{eqnarray}

\begin{eqnarray}
\omega&=&\left(-\frac{3}{2g^4}-\frac{4}{g^2}+8\right)\arctan\left(\frac{1}{
2g}\right)\nonumber\\
&+&\left(\frac{3}{2g^4}+\frac{3}{g^2}-\frac{5}{2}\right)\arctan\left(
\frac{1-g^2}{2g}\right)\nonumber\\
&+&\frac{\pi}{2g^2}-\frac{5\pi}{4}+\frac{1}{g}.
\end{eqnarray}
The function $\omega$ and the anomalous dimension $\eta$ have well defined 
limits as $g\to 0$, which correspond to the limit of the conventional GL 
model. As $g\to 0$, we find $\eta\to -f/4$ as $g\to 0$ while 
$\omega\to 3\pi/2$. The opposite limit, $g\to\infty$, leads to the expected 
decoupling of the gauge and scalar degrees of freedom since both $\eta$ and 
$\omega$ tend to zero in this limit. The $f^2$ term 
in Eq.(\ref{u}) comes from 
the graph shown in Fig.1. As mentioned in the introduction, this graph 
is zero when $p=0$. In the scaling considered in this paper this graph 
survives since $p\neq 0$ and we can recover the $g\to 0$ limit.

The charged fixed points are given by $f^{*}=16$, $g^{*}$ arbitrary and 

\begin{equation}
u^{*}_{\pm}=\frac{4}{5}\left[1-2\eta\pm\sqrt{(2\eta-1)^2-\frac{160}{
\pi}\omega^{*}}\right],
\end{equation}
where $\eta=\eta_{\psi}(f^{*},g^{*})$ and $\omega^{*}=\omega(g^{*})$. 
We have that $u^{*}_{\pm}$ is real only 
for $g^{*}\geq g^{*}_{c_{1}}\approx 0.42$, 
which corresponds to the condition for the existence of charged fixed 
points in our model. The flow diagram is shown in Fig.2 for $g^{*}=0.48$. 
The left charged fixed point, coresponding to $u^{*}_{-}$, is the so called 
tricritical fixed point. The tricritical fixed point is attractive over a 
line intercepting the origin called the tricritical line. The 
tricritical line separates the regions of first and second order phase 
transitions. The right charged fixed point, corresponding to $u^{*}_{+}$, is 
infrared attractive and describes the physics of second order phase 
transitions in superconductors. Note also in the flow diagram the Gaussian 
and the $XY$ fixed points. Another interesting point is that there exists
another critical value of $g^{*}$, $g^{*}_{c_{2}}\approx 0.81$, such that 
for $g^{*}>g^{*}_{c_{2}}$ the tricritical fixed point is in the region of 
the plane $uf$ defined by $u<0$. In this region the 
tricritical fixed point lost its physical meaning. It results that the only 
charged fixed point is the infrared one. Consequently, for 
$g^{*}>g^{*}_{c_{2}}$ only second order behavior seems to be possible.

Let us evaluate the critical exponents for the superconducting transition in 
function of $g*>g^{*}_{c_{1}}$. It is sufficient to evaluate $\eta$ and $\nu$ 
since the other exponents are obtained from these ones via scaling relations. 
$\eta$ is given simply by evaluating $\eta_{\psi}$ at the fixed point. The 
evaluation of $\nu$, however, is more involved. In the critical theory this 
exponent is evaluated by considering a $\psi^{\dag}\psi$ insertion in the 
two point function. Thus, we must compute the 1-particle irreducible function 
$\Gamma_{\psi}^{(1,2)}$ subjected to the renormalization condition:

\begin{equation}
\Gamma_{\psi}^{(1,2)}(k_{1},k_{2};-k_{1}-k_{2})|_{S.P.}=1,
\end{equation} 
which will determine the renormalization constant of $\psi^{\dag}\psi$, 
denoted by $Z_{\psi}^{(2)}$. As usual \cite{ZJ}, the $\nu$ exponent is 
given by the fixed point value of the RG function $\nu_{\psi}$ defined by

\begin{equation}
\label{nu}
\frac{1}{\nu_{\psi}}=2+p\frac{d}{dp}\log\left(\frac{Z^{(2)}_{\psi}}{
Z_{\psi}}\right).  
\end{equation}
We find:

\begin{eqnarray}
\eta_{\psi}^{(2)}&\equiv&p\frac{d\log Z_{\psi}^{(2)}}{dp}
\nonumber\\
&=&-\frac{u}{4}
-\frac{f}{4\pi}\left[\frac{(3-4g^2)(3+4g^2)}{8\Delta^{3/2}}\arctan\left(
\frac{\sqrt{\Delta}}{g}\right)\right.\nonumber\\
&+&\frac{3+4g^2}{2\sqrt{3}g^2}\arctan\left(\frac{3-4g^2}{4\sqrt{3}g}
\right)-\pi\frac{3-4g^2}{4\sqrt{3}g^2}\nonumber\\
&+&\left.\frac{(3-4g^2)(3+4g^2)}{8g\Delta}\right],
\end{eqnarray}
where

\begin{equation}
\Delta=g^4+\frac{g^2}{2}+\frac{9}{16}.
\end{equation}
At 1-loop order we obtain

\begin{equation}
\nu_{\psi}\approx\frac{1}{2}\left(1-\frac{1}{2}\eta_{\psi}^{(2)}+\frac{1}{2}
\eta_{\psi}\right).
\end{equation}
The Fig.3 shows a plot of the exponent $\nu$ as a function of $g^{*}$. Note 
that the plot is made for $g^{*}\geq g^{*}_{c_{1}}$ which corresponds to the 
region where charged fixed points should exist. We observe that as $g^{*}$ 
increases the value of $\nu$ tends asymptotically to $0.6$, which 
corresponds approximately to the XY value for the pure scalar model 
in the 1-loop approximation at 
fixed dimension. This is an expected result since for $g\to\infty$ the 
gauge modes decouple from the scalar modes. This result can be verified 
directly from the above RG functions by expanding for $g$ large.

In recent years it has been stablished that the $\nu$ exponent for the 
(non-topological) superconducting transition is given by 
$\nu=\nu_{XY}\approx 0.67$ \cite{Lawrie,KleinertI,HerbutI}. Thus, we are 
tempted to search for what value of $g^{*}$ we have $\nu=\nu_{XY}$. One 
obtains that $\nu\approx 0.67$ for $g^{*}\approx 0.776$. This is smaller than 
$g^{*}_{c_{2}}$ and we have still the region of first order transition in 
the flow diagram. Unfortunately, a pathological behavior arises for this 
value of $g^{*}$. The trouble comes from the $\eta$ exponent. Indeed, we 
have that $\eta\approx-2.47<-1$ while we know from the scaling relations 
that the condition $\eta>-1$ must be fulfilled. This result is probably an 
artifact of our 1-loop perturbation theory. Another related
problem is that the 
above results fail in describing a flow diagram with charged fixed points 
in the limit $g\to 0$. In this limit our flow diagram resembles to that 
one obtained by Halperin {\it et al.} in the seventies \cite{HLM} and 
, as mentioned in the introduction, their result is an artifact of 
the $\epsilon$-expansion. Therefore, we should improve 
our perturbation theory in order to bypass all these difficulties.  

\section{Improved RG results}

As we have said in the last section, the beta functions for the topological 
GL model have a well defined $g\to 0$ limit. However, we have seen that in 
this limit no charged fixed points exist. Moreover, the $\eta$ exponent 
could attain unphysical values which violate the scaling relations.   
The trouble is that there are in fact two fundamental lenght scales in 
this problem, namely, the correlation lenght $\xi$ and the magnetic field 
penetration depth, $\lambda$. The $\xi$ is related to the scaling of the 
scalar field while $\lambda$ is related to the scaling of the gauge field. 
Thus, in principle, the renormalization conditions for the gauge coupling 
should be fixed at a different point of the scalar coupling. Since there is  
a relation between the lenghts $\xi$ and $\lambda$, we must have 
also a relation between the corresponding renormalization points. We  
develop this more general point of view by using a simple method    
suggested recently by Herbut and Tesanovic \cite{Herbut}. It consists 
in fixing the renormalization condition Eq.(6) at the point given by 
$k^2=p^2/c^2$, $c$ giving in this way the ratio between the two scales of 
the problem. 
If we use the same reasoning here 
we obtain the following beta functions in the $g\to 0$ limit:

\begin{eqnarray}
\label{ff}
p\frac{df}{dp}&=&-f+\frac{cf^2}{16},\\
p\frac{du}{dp}&=&(2\eta_{\psi}-1)u+\frac{5}{8}u^2+\frac{3}{8}f^2,
\end{eqnarray}
where $\eta_{\psi}=-f/4$. Note that we have not exactly the same beta 
functions as in ref. \cite{Herbut}. This is due to the fact that we used 
the convention that all external momenta are incoming at the vertices. Our 
choice is more usual in the field theoretical literature and has the 
advantage that corresponding crossed graphs have the same value at the 
symmetrical point. However, these differences in the conventions does 
not change appreciably the physical results like the values of the 
critical exponents.

The value of $c$ can be fixed by demanding that, if charged fixed points 
do exist, it should happen at the same critical scale. 
At 1-loop this is equivalent 
to demanding that the Ginzburg parameter $\kappa=u/2f$ should be invariant 
along the RG trajectory connecting the origin and the tricritical fixed 
point \cite{Herbut}. 
We choose the tricritical fixed point because there are good 
numerical estimates of $\kappa$ available \cite{Bart,KleinertII}. If 
$\kappa_{tric}$ denotes the value of $\kappa$ at the tricritical fixed 
point we obtain that c is determined by the following equation:

\begin{equation}
\label{c}
\kappa_{tric}^{2}=\frac{c+8-\sqrt{c^2+16c-176}}{40}.
\end{equation}
We take $\kappa_{tric}\approx 0.42/\sqrt{2}$ which is the value obtained 
from Montecarlo calculations \cite{Bart}. Thus, Eq.(\ref{c}) gives 
$c\approx 27.78$. For this value of $c$ we have the flow diagram shown 
in Fig.4 for the non-topological model. Fig. 5 shows the detail of the 
region near the tricritical line which is not well seen in Fig. 4.

The RG function $\nu_{\psi}$ in the $g\to 0$ limit is given by

\begin{equation}
\nu_{\psi}\approx\frac{1}{2}\left[1+\frac{u}{8}-\frac{f}{4}\left(\frac{1}{2}
-\frac{2+\sqrt{3}}{3}\right)\right]
\end{equation}
We have the following result for the critical exponents:

\begin{eqnarray}
\nu&\approx&0.676,\\
\eta&\approx&-0.14,
\end{eqnarray}
a result in good agreement with the expected $XY$ behavior. In ref. 
\cite{Herbut} the value $\nu\approx 0.53$ was obtained by using a $c$ 
corresponding to $\kappa_{tric}=0.8/\sqrt{2}$ which is determined from the 
lattice dual model \cite{KleinertII}. However, they obtained 
$\nu\approx 0.62$ for $\kappa_{tric}=0.42/\sqrt{2}$. 
It is possible to improve this value obtained in ref.\cite{Herbut} by 
using directly Eq.(\ref{nu}) without expanding it, to obtain the better 
value $\nu\approx 0.67$, in close agreement with our result \cite{Note2}. This 
$XY$ value has also be obtained in ref.\cite{KleinertI}.
It is worth to mention 
that non-perturbative RG calculations by Bergerhoff {\it et al.} 
\cite{Berg} gives $\nu\approx 0.53$ in a certain approximation 
corresponding to a truncation of the average action (the Legendre transform 
of the Wilsonian effective action) at $|\psi|^4$. Truncation at 
$|\psi|^8$ gives the improved result $\nu\approx 0.58$.

Let us come back now to the case of non-vanishing $g$.  
Since the $g$ coupling is associated to the 
gauge degrees of freedom its momentum scale  
should be rescaled by $c$. Thus, in addition to Eqs.(\ref{u}) 
and (\ref{ff}) we have   

\begin{equation}
p\frac{dg}{dp}=g\left(-1+\frac{cf}{16}\right).
\end{equation}
We keep the same value of $c$ in order to preserve the features 
of the $g\to 0$ limit. Now we have charged fixed points even when $g=0$ and 
therefore there is no $g^{*}_{c_{1}}$. On the other hand, we 
have still $g_{c_{2}}^{*}$. We verified that $g^{*}_{c_{2}}$ 
is the same as before, that is, $g^{*}_{c_{2}}\approx 0.81$. This means 
that for $g^{*}>0.81$ we have only second order behavior.
In Fig. 6 we show the flow diagram for $g^{*}=0.48$ with the detail of the 
tricritical region shown in Fig. 7. In Fig. 8 we plot the exponent $\nu$ as 
a function of $g^{*}$ in the improved 1-loop calculation. Note that the shape 
of the plot is qualitatively the same as in Fig.3. The main 
difference is that the vertical scale 
is compressed and the most remarkable feature is the fact that $\nu$ is now 
much less sensible to the value of $g^{*}$, showing only a small deviation 
from the $XY$ scaling.
For instance, the exponents 
$\nu$ and $\eta$ for $g^{*}=0.48$ are given by

\begin{eqnarray}
\nu&\approx&0.628,\\
\eta&\approx&-0.11.
\end{eqnarray} 
Of course, we have still the same $g\to\infty$ limit with exponents 
$\eta=0$ and $\nu=0.6$, corresponding to the decoupled situation.

\section{Discussion}

The main aim of this paper is to initiate a careful study of a topological 
GL model from the point of view of critical phenomena. For this reason we 
have concentrated the efforts on the RG flow and the evaluation of critical 
exponents. The results show that the topological coupling is a good 
control parameter with respect to the fixed point structure. For instance, 
we have seen that the region of first order transition is crunched as the 
topological coupling is increased and eventually the type I behavior is 
lost. An interesting point is that the $\nu$ exponent does not fluctuate 
very much around the $XY$ value. 

Besides superconductivity, the topological GL model may be useful in other 
physical contexts. For example, it can be applied in the study of soft 
materials like the chiral liquid crystals \cite{Lubensky}. In this case 
the gauge field should be thought as a director field and the CS term is 
used in order to introduce the chirality for the constituting molecules.

Another interesting problem is the physics of the chiral spin state. This 
state arises when one considers the Hamiltonian for the Heisenberg 
antiferromagnet whose spins interact not only through nearest neighbors 
interaction, but also through a next-nearest neighbor one \cite{Wen,Fradkin}. 
The next-nearest neighbor interaction frustrate the N\'eel state and 
generates a mean field solution corresponding to a disordered spin state. 
The most stable configuration corresponds to the so called chiral spin 
state. The continuum effective theory is obtained by computing the 
fluctuations around such a mean field ground state and it contains a 
dynamically generated gauge field. The gauge sector of the effective action 
is identical to that one of the topological GL model. In this case the 
gauge couplings are given in terms of mean field parameters of the 
original model (for details see ref.\cite{Fradkin}).

Finally, we hope that this work will contribute to improve the 
understanding of GL models. 

\section*{Ackowlegments}

C. de Calan would like to thank the hospitality of the Centro Brasileiro 
de Pesquisas F\'{\i}sicas (CBPF) where part of this work has been done. 
F. S. Nogueira would like to thank the financial support of the agency 
CNPq, a division of the Brazilian Ministry of Science and Technology.

\begin{figure}
\centerline{\psfig{figure=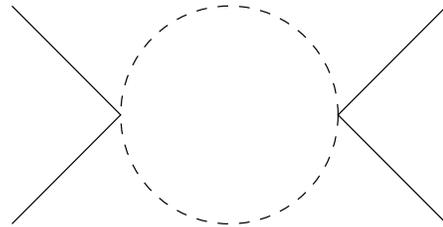,height=3truecm,angle=-90}}
\caption{Graph contributing to the $f^2$ term in the beta function for u. 
The dashed lines represent the gauge field propagator.}
\end{figure}

\begin{figure}
\centerline{\psfig{figure=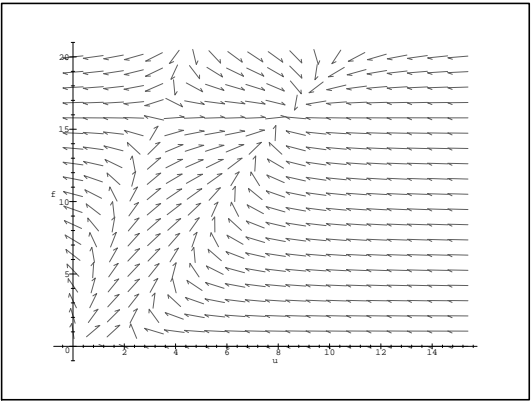,height=6truecm,angle=-90}}
\caption{Flow diagram for $g^{*}=0.48$.}
\end{figure}

\begin{figure}
\centerline{\psfig{figure=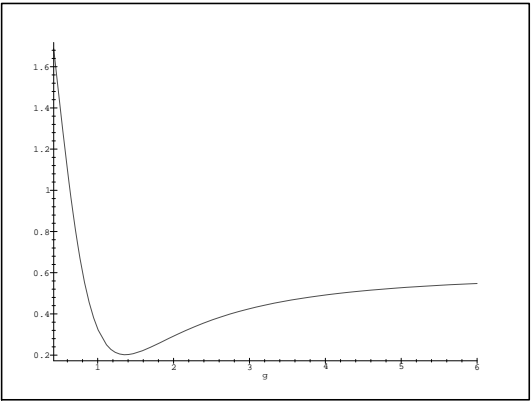,height=6truecm,angle=-90}}
\caption{Plot of $\nu$ as a function of $g^{*}$ for 
$g^{*}\geq 0.42$}
\end{figure}

\begin{figure}
\centerline{\psfig{figure=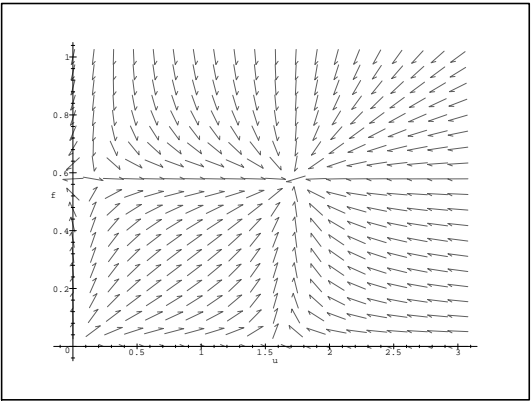,height=6truecm,angle=-90}}
\caption{Flow diagram for the non-topological model.}
\end{figure}
\begin{figure}
\centerline{\psfig{figure=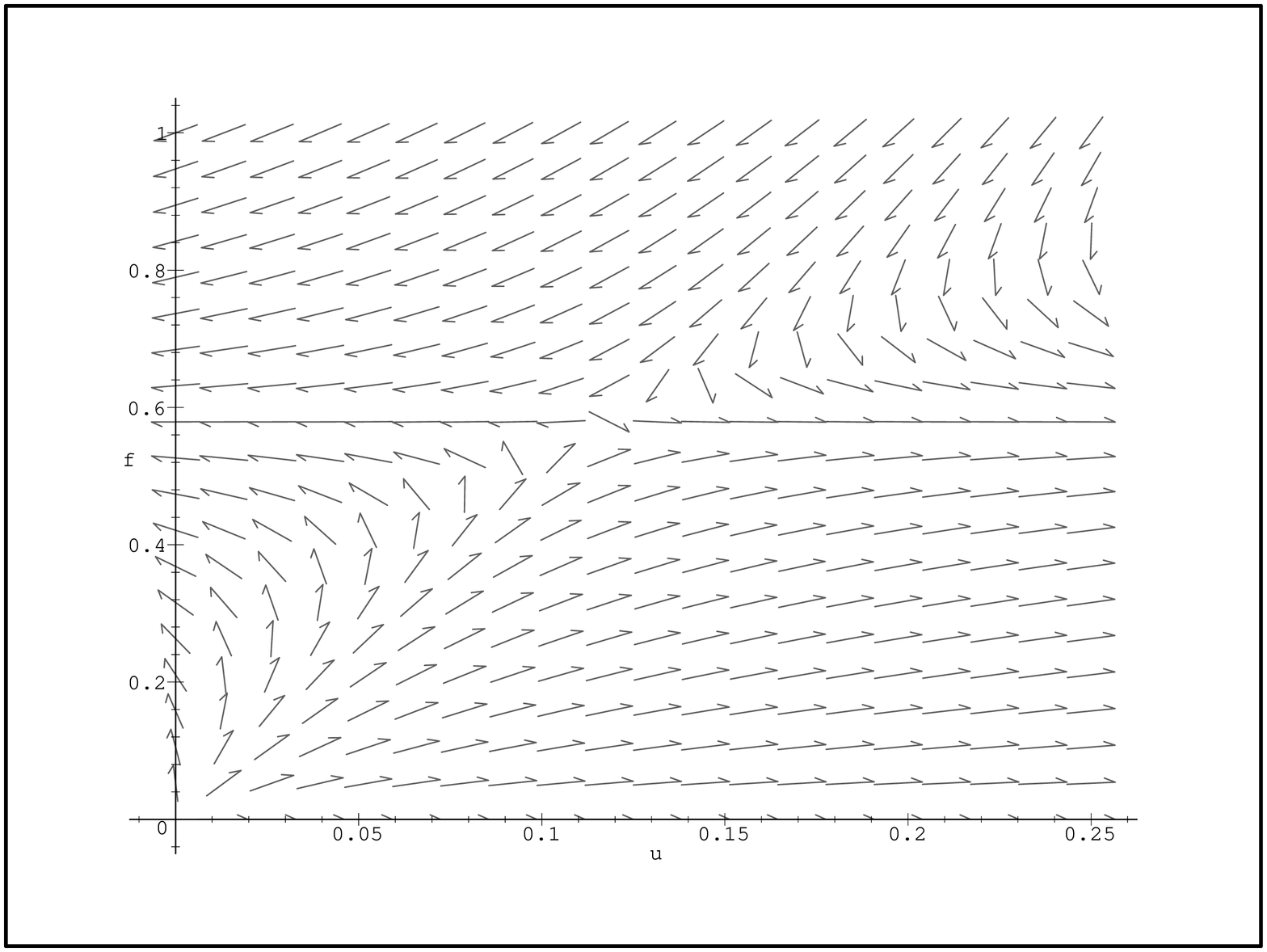,height=6truecm,angle=-90}}
\caption{Detail of Fig.4 in the region near the tricritical line.}
\end{figure}

\begin{figure}
\centerline{\psfig{figure=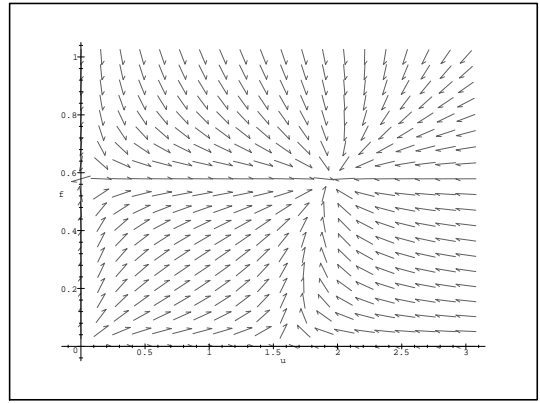,height=6truecm,angle=-90}}
\caption{Flow diagram for the topological model with improved 1-loop 
calculations for $g^{*}=0.48$.}
\end{figure} 
\begin{figure}
\centerline{\psfig{figure=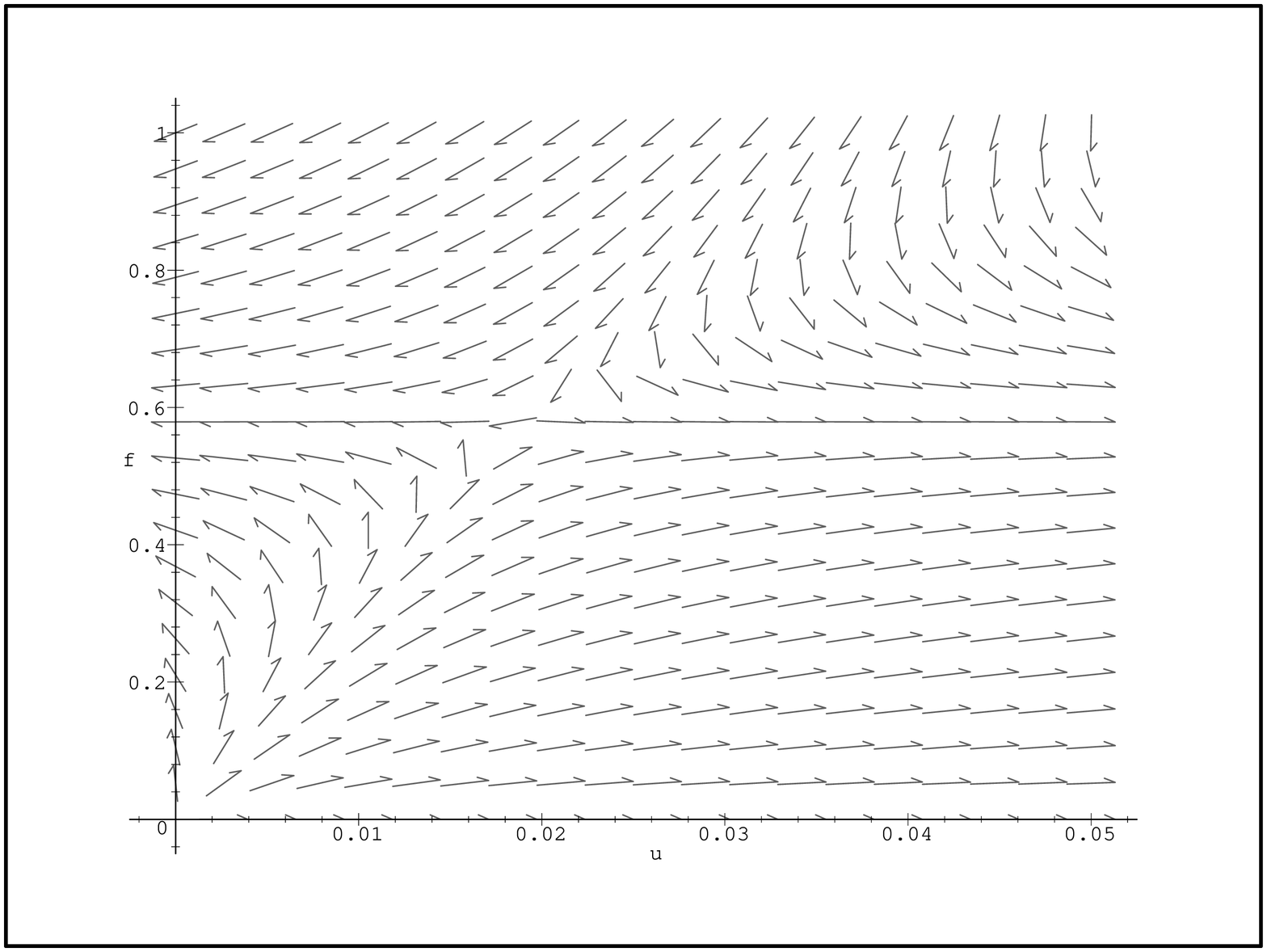,height=6truecm,angle=-90}}
\caption{Detail of Fig.6 in the region near the tricritical line.}
\end{figure}
\begin{figure}
\centerline{\psfig{figure=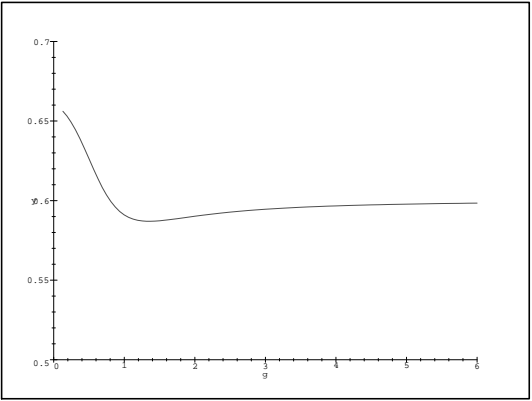,height=6truecm,angle=-90}}
\caption{Plot of the exponent $\nu$ as a function of $g^{*}$.}
\end{figure}

\end{document}